\documentclass[]{aa}
\usepackage{graphicx}
\usepackage{txfonts}
\usepackage{natbib}
\bibpunct{(}{)}{;}{a}{}{,}

\begin{document}

\title{Formation of a solar H$\alpha$ filament from orphan penumbrae}
%\titlerunning{}

\author{D. Buehler \inst{1}, A. Lagg \inst{1}, M. van Noort \inst{1} \and  S.K. Solanki \inst{1}$^,$\inst{2}}
\authorrunning{D. Buehler, et al.}

\institute{\inst{1} Max Planck Institute for Solar System Research, Justus-von-Liebig-Weg 3, 37077 G\"ottingen, Germany\\
\inst{2} School of Space Research, Kyung Hee University, Yongin, Gyeonggi, 446-701, Korea}

\date{Received 15 October 2015 / Accepted 16 March 2016}

\abstract
{}
{The formation and evolution of an H$\alpha$ filament in active region (AR) 10953 is described.}
{Observations from the Solar Optical Telescope (SOT) aboard the Hinode satellite starting from UT 18:09 on $27^{th}$ April 2007 until UT 06:08 on $1^{st}$ May 2007 were analysed. 20 scans of the $6302\AA$ Fe I line pair recorded by SOT/SP were inverted using the spatially coupled version of the SPINOR code. The inversions were analysed together with co-spatial SOT/BFI G-band and Ca II H and SOT/NFI H$\alpha$ observations.}
{Following the disappearance of an initial H$\alpha$ filament aligned along the polarity inversion line (PIL) of the AR, a new H$\alpha$ filament formed in its place some 20 hours later, which remained stable for, at least, another 1.5 days. The creation of the new H$\alpha$ filament was driven by the ascent of horizontal magnetic fields from the photosphere into the chromosphere at three separate locations along the PIL. The magnetic fields at two of these locations were situated directly underneath the initial H$\alpha$ filament and formed orphan penumbrae already aligned along the H$\alpha$ filament channel. The 700 G orphan penumbrae were stable and trapped in the photosphere until the disappearance of the overlying initial H$\alpha$ filament, after which they started to ascend into the chromosphere at $10\pm5$ m/s. Each ascent was associated with a simultaneous magnetic flux reduction of up to $50\%$ in the photosphere. The ascended orphan penumbrae formed dark $\emph{seed}$ structures in H$\alpha$ in parallel with the PIL, which elongated and merged to form an H$\alpha$ filament. The filament channel featured horizontal magnetic fields of on average 260 G at $\log(\tau)=-2$ suspended above the nearly field-free lower photosphere. The fields took on an overall $\emph{inverse}$ configuration at $\log(\tau)=-2$ suggesting a flux rope topology for the new H$\alpha$ filament. The destruction of the initial H$\alpha$ filament was likely caused by the flux emergence at the third location along the PIL.}
{ We present a new interpretation of the H$\alpha$ filament formation in AR 10953 whereby the mainly horizontal fields of orphan penumbrae, aligned along the H$\alpha$ filament channel, ascend into the chromosphere, forming $\emph{seed}$ fragments for a new, second H$\alpha$ filament. The orphan penumbral fields ascend into the chromosphere  $\sim$9-24 hours before the H$\alpha$ filament is fully formed.}


\maketitle

\section{Introduction}
Solar H$\alpha$ filaments are among the longest \citep{deslandres1894,waldmeier1938,zirin1960} and best studied \citep[see reviews by][]{hirayama1985,tandberg1995,mackay2010} objects in the Sun's atmosphere. They are easily identified as dark filaments on the disc \citep[e.g.][]{lin2005} or as prominences suspended above the solar limb \citep[e.g.][]{okamoto2007} when the Sun is observed in a strong spectral line such as H$\alpha$. The comparatively cool, dense material making up the bulk of the H$\alpha$ filament is prevented from immediately returning to the photosphere by magnetic fields pervading it.

Photospheric magnetograms revealed that H$\alpha$ filaments are typically located along a polarity inversion line (PIL) \citep{babcock1955} flanked by kG magnetic field concentrations (MFCs). Running along the long axis of H$\alpha$ filaments are typically hG horizontal fields \citep{leroy1983,leroy1984,kuckein2009,kuckein2012,xu2012}. Beginning with the earliest H$\alpha$ filament models \citep{kippenhahn1957}, the H$\alpha$ filament's material is proposed to reside in dips in the magnetic field connecting the opposite polarities on either side of the PIL. The model by \citet{kippenhahn1957} and its various extensions by \citet{malherbe1983,heinzel2001} proposes a $\emph{normal}$ polarity configuration, whereby the magnetic field lines connect the opposite polarities flanking the PIL akin to a potential field configuration. The $\emph{inverse}$ or O-loop configuration, suggested by \citet{kuperus1967} \& \citet{kuperus1974}, but see also \citet{priest1989,ballegooijen1989} tends to be the favoured configuration given the variety of observations supporting it \citep[e.g.][]{bommier1994,lites1995,dere1999,lopez2006,okamoto2008,kuckein2012}, although observations in favour of the \citep{kippenhahn1957} model have been reported by \citet{leroy1984}. However, complicated internal motions observed within H$\alpha$ filaments appear to call for even more complicated models \citep[e.g.][]{schmieder1991,berger2008,sasso2011}. 

The formation process of H$\alpha$ filaments is less well understood. A large number of theoretical models of this process exist. The $\emph{surface}$ models typically involve atmospheric reconnection of already emerged fields brought about by shear flows and/or inflows along the PIL \citep[e.g.][]{ballegooijen1989,devore2005}. Alternatively, $\emph{sub-surface}$ models employ the emergence of new, twisted flux ropes through the photosphere and into the chromosphere to form a new H$\alpha$ filament \citep[e.g.][]{rust1994,low1995}. A more complete overview on these models is given in \citet{mackay2010}.

Direct observations of H$\alpha$ filament creation are comparatively rare. \citet{gaizauskas1997,gaizauskas2001} observed inflow motions of MFCs in magnetograms into a PIL over several days and \citet{wang2007} observed shearing in H$\alpha$ fibrils, leading to H$\alpha$ filament creation and supporting the $\emph{surface}$-type models. In contrast, \citet{lites1995} and \citet{okamoto2008,okamoto2009} observed the emergence of a flux rope along a PIL also leading to H$\alpha$ filament creation and supporting the $\emph{sub-surface}$ models. However, the H$\alpha$ filament creation observed by \citet{lites1995} took place in conjunction with $\emph{delta}$ sunspots, which are rarer than H$\alpha$ filaments, whilst the flux emergence observed by \citet{okamoto2008,okamoto2009} failed to show any of the typical flux emergence signs \citep{cheung2014}. The investigations by \citet{kuckein2012vel} \& \citet{xu2012}, whilst generally supporting the rising flux rope scenario, lacked the required temporal resolution to unambiguously characterise the evolution of their magnetic structures. The observations by \citet{gaizauskas1997,gaizauskas2001,wang2007} employed low resolution data ($1''-2''$) and the necessary shear flows along the PIL have so far never been directly observed during H$\alpha$ filament creation.

Given the need for further and more detailed observations on the creation of H$\alpha$ filaments, we analysed active region (AR) 10953, which hosted two consecutive, well studied H$\alpha$ filaments \citep{okamoto2008,okamoto2009,wheatland2009,su2009,canou2010}. The description of the evolution and the formation process of the second H$\alpha$ filament is the aim of this paper.

\section{Observations \& Analysis}
This investigation employs data sets recorded by the narrow-band filter imager (NFI), the broad-band filter imager (BFI) and the spectropolarimeter (SP), all of which are part of the Solar Optical Telescope (SOT) \citep{tsuneta2008sot,suematsu2008sot,ichimoto2008sot,shimizu2008sot} aboard the Hinode satellite \citep{kosugi2007}. The SOT was extensively used to monitor the photospheric and chromospheric evolution of AR 10953 at a high spatial and spectral resolution. The AR was tracked from $27^{th}$ April 2007 until  $7^{th}$ May 2007, which encompassed the AR's passage across the solar disc. The PIL of the AR hosted successive H$\alpha$ filaments. Unfortunately, the pointing of the SOT was such that beginning at the $1^{st}$ of May the H$\alpha$ filament was largely outside the field of view (FOV) so that data sets of the AR recorded afterwards are not included in this investigation.

The SOT/SP records two magnetically sensitive photospheric Fe I lines at 6302 $\AA$ with a spectral resolution of 30 $m\AA$. At each slit position the four Stokes parameters $I$, $Q$, $U$ and $V$ are recorded. The majority of the observations were performed using the $\emph{fast}$ mode, which is characterised by an exposure time of 1.6 s per slit position and effective pixels size of $0\farcs32$. Three data sets at UT 09:00 \& 15:00 on $28^{th}$ April and UT 18:35 on $30^{th}$ April were obtained in the $\emph{normal}$ mode, which has an exposure time of 4.8 s per slit position and effective pixel size of $0\farcs16$. Due to the different spatial sampling and exposure times both observation modes have a noise level of $1 \times 10^{-3}$ $I_c$. All the SOT/SP data sets used in this investigation are listed in Tab. \ref{SPtable} and are typically 3 - 5 hours apart from each other. Each data set was reduced using $\emph{sp\_prep}$ \citep{lites2013} and then inverted using the SPINOR code \citep{frutiger2000}.

The inversion procedure follows the 2D spatially coupled inversion technique introduced by \citet{vannoort2012}, which has subsequently been applied to many SOT/SP data sets \citep{riethmueller2013,vannoort2013,tiwari2013,lagg2014,buehler2015}. The technique takes into account the local image degradation caused by SOT/SP's point-spread-function (PSF) during the inversion procedure itself. Since the global straylight contribution in SOT/SP observations is below $10\%$ \citep{danilovic2008,vannoort2012}, the central part of the PSF accounts for the bulk of the expected rms contrast in the observations.  As a result the observed spectra can be successfully described in terms of a single atmosphere per pixel and the need to introduce a filling or straylight factor is removed.  Spectra of the photosphere routinely display both amplitude and area asymmetries in their Stokes parameters \citep{solanki1993,stenflo2010,viticchie2010}, which are indicative of line-of-sight (LOS) gradients in the atmosphere. The fitted model atmospheres of the inversion could account for these LOS gradients through the use of three nodes in optical depth, at $\log(\tau)= 0, -0.8$ $\& -2$. At each node the atmospheric parameters, the temperature, $T$, the magnetic field, $B$, the LOS field inclination, $\gamma$, the LOS field azimuth, $\phi$, the LOS velocity, $v$, and microturbulence, $\xi$, could be altered. Values belonging to optical depths other than the three nodes were obtained from spline interpolations through the three nodes. By solving the radiative transfer equation using the STOPRO routines \citep{solanki1987phd}, which are part of the SPINOR code, synthetic Stokes spectra of a given model atmosphere can be generated and compared to observed spectra. Through the use of response functions and a Levenberg-Marquardt algorithm, which minimises a $\chi^2$ merit function, the atmospheric parameters of a pixel's model atmosphere can be iteratively fitted until the synthesised spectra closely match the observed ones.

All the LOS inclinations and LOS azimuths obtained from the inversions were subsequently converted to local solar coordinates and the $180^{\circ}$ azimuth ambiguity was removed. The azimuth ambiguity removal routine follows the principles outlined in \citet{buehler2015} whereby the magnetic orientation of canopy pixels at $\log(\tau)=-2$ of a MFC is resolved by relating them to the core pixels of the MFC under the assumption that the magnetic field is approximately divergence free. The remaining pixels are then resolved by local dot products with already azimuth resolved pixels. In regions with high magnetic field strengths such as a sunspot the $j_{z}$ component of the current is additionally taken into account. Pixel with field strengths below 50 G were not resolved.  The corrected inclinations and azimuths are labelled $\Gamma$ and $\Phi$ respectively. The zero velocity was set by forcing the mean umbral LOS velocity to zero, which required a correction of $230\pm50$ m/s for the various SOT/SP data sets.
 
\begin{table}
\caption{Analysed SOT/SP scans.}              % title of Table
\label{SPtable}      % is used to refer this table in the text
\centering                                      % used for centering table
\begin{tabular}{c c c c}          % centered columns (4 columns)
\hline\hline                        % inserts double horizontal lines
\bf{No} & \bf{Time [UT]} & \bf{$X$ $['']$} & $\mu$\\    % table heading
 & 2007 April/May & $Y\sim-100''$\\
\hline                                   % inserts single horizontal line
    1 & 27 -- 18:09 & -895 & 0.32\\      % inserting body of the table
    \bf{2} & 28 -- 09:00 & -731 & 0.63\\
    3 & 28 -- 11:40 & -716 & 0.65\\
    \bf{4} & 28 -- 15:00 & -691 & 0.68\\
    5 & 28 -- 18:14 & -670 & 0.70\\
    6 & 28 -- 21:10 & -649 & 0.72\\
    7 & 29 -- 00:17 & -626 & 0.74\\
    8 & 29 -- 03:30 & -602 & 0.77\\
    9 & 29 -- 08:00 & -568 & 0.79\\
   10& 29 -- 11:27 & -541 & 0.82\\
   11& 29 -- 15:20 & -510 & 0.84\\
   12& 29 -- 20:00 & -471 & 0.86\\
   13& 30 -- 01:00 & -430 & 0.89\\
   14& 30 -- 04:51 & -397 & 0.90\\
   15& 30 -- 08:40 & -364 & 0.92\\
   16& 30 -- 12:30 & -341 & 0.93\\
   17& 30 -- 15:30 & -315 & 0.94\\
   \bf{18}& 30 -- 18:35 & -299 & 0.94\\
   19& 30 -- 22:30 & -253 & 0.96\\
   20& 01 -- 01:50 & -223 & 0.97\\
\hline                                             %inserts single line
\end{tabular}
\tablefoot{\emph{Normal} mode scans are marked in \emph{bold}.}
\end{table}
The spatial resolution of $\emph{fast}$ mode SOT/SP images is only $0\farcs64$, while an individual flux tube in the form of a bright point can have a width of only $0\farcs15$ \citep{lagg2010} or even less \citep{riethmueller2014}. As a result, the magnetic field strength values obtained from the inversions should be viewed as a lower limit of the true values, in particular for smaller magnetic structures. However, this investigation focusses on magnetic fields involved in the evolution of an H$\alpha$ filament surrounded by an extensive plage/pore region as well as a sunspot, all of which are structures which are generally several times the spatial resolution limit of $0\farcs64$ in size. Any fine structure belonging to these magnetic fields is, nonetheless, largely lost and will only be lightly touched upon in the following sections. Finally, the inversion results support and complement the co-temporal and co-spatial SOT/BFI and SOT/NFI observations of the AR.

The BFI performed observations of the G-band at 4305 $\AA$, which forms in the photosphere, and Ca II H at 3968 $\AA$, which mainly depicts the upper photosphere due to the broadness of the employed filter \citep{carlsson2007}. The BFI observations possess the highest angular resolution at $0\farcs22$. The NFI observed the AR in the H$\alpha$ line at 6563 $\AA$ with an angular resolution of $0\farcs32$. The H$\alpha$ observations suffer from an air bubble, which obscures and distorts part of the image and places an observable intensity pattern across the remaining image. Fortunately, the H$\alpha$ filament investigated in this study was located far from the air bubble, allowing its evolution to be monitored continuously. All three channels performed observations of the AR with a cadence of one image per minute starting at UT 11:39 on $28^{th}$ April 2007 until UT 06:07 on $1^{st}$ May 2007, except for small gaps of 15 minutes every 2 - 3 hours. All three channels were reduced using $\emph{fg\_prep}$. The observations overlap with the SOT/SP scans save for the initial two in Tab. \ref{SPtable}. The BFI, NFI and SP observations were aligned to each other using a rigid alignment routine available from the $\emph{solarsoft}$ package.   

\section{Results}
\begin{figure*}
       \centering
        \includegraphics[width=18cm]{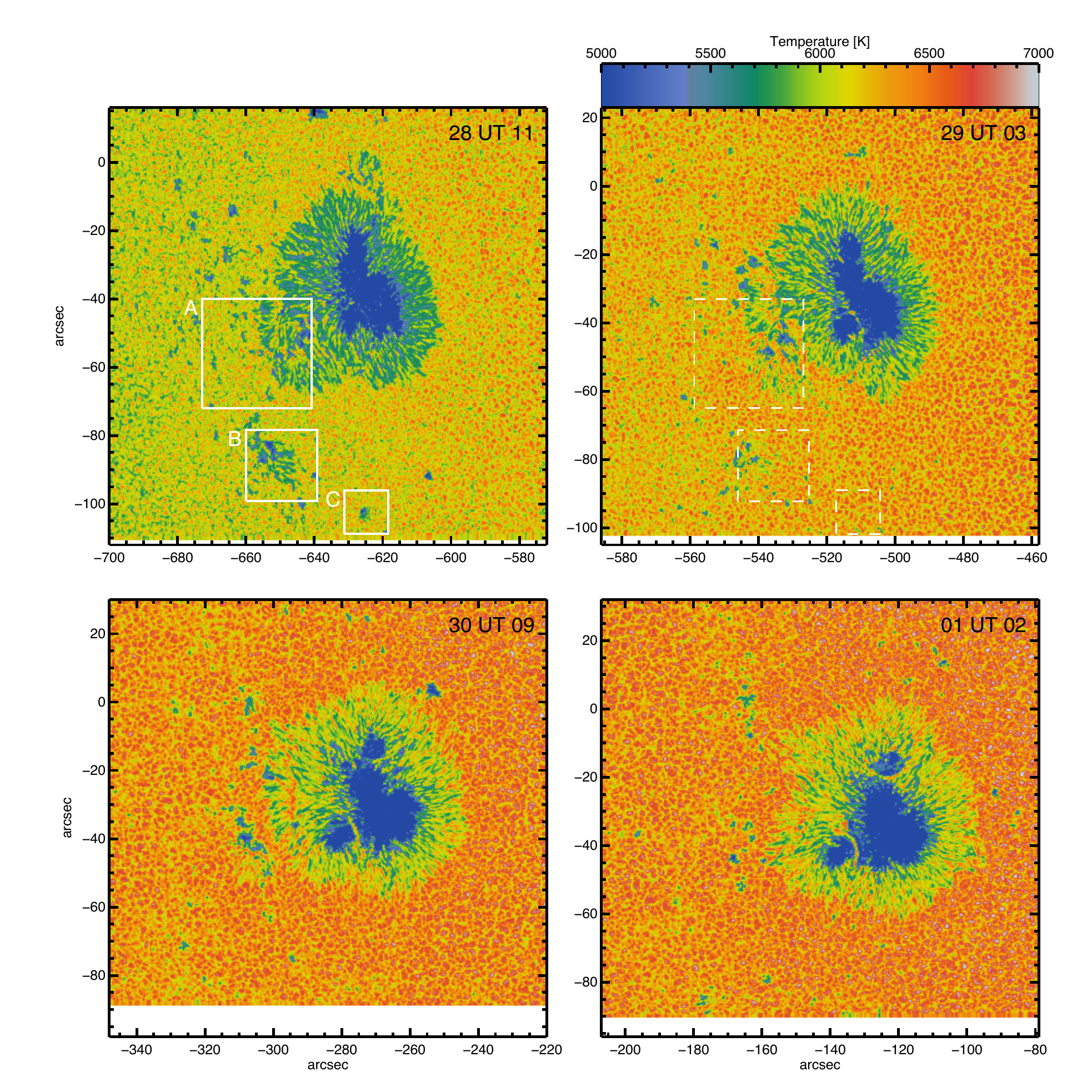}
        \caption{Temperatures at $\log(\tau)=0$ from the inversion of four SOT/SP scans. The colour bar applies to all plots. The $xy$ axes indicate the distance to disc centre. The three boxes marked $A$, $B$ $\&$ $C$ enclose regions of rising flux.}
         \label{Overviewevo}
         \end{figure*}

A general impression of AR 10953's evolution can be gained from Fig. \ref{Overviewevo}, which depicts the temperature at $\log(\tau)=0$ of four different SOT/SP scans. At UT 11 on $28{th}$ April the AR harbours a fully developed sunspot surrounded by pores and orphan penumbrae. By UT 02 on $1^{st}$ May many of these surrounding structures have disappeared and the AR appears to have lost some of its complexity. Three regions, enclosed by the boxes $A$, $B$ and $C$, undergo the greatest change in time and play a critical role in the formation of the AR's H$\alpha$ filament.

\begin{figure*}
       \centering
        \includegraphics[width=18cm]{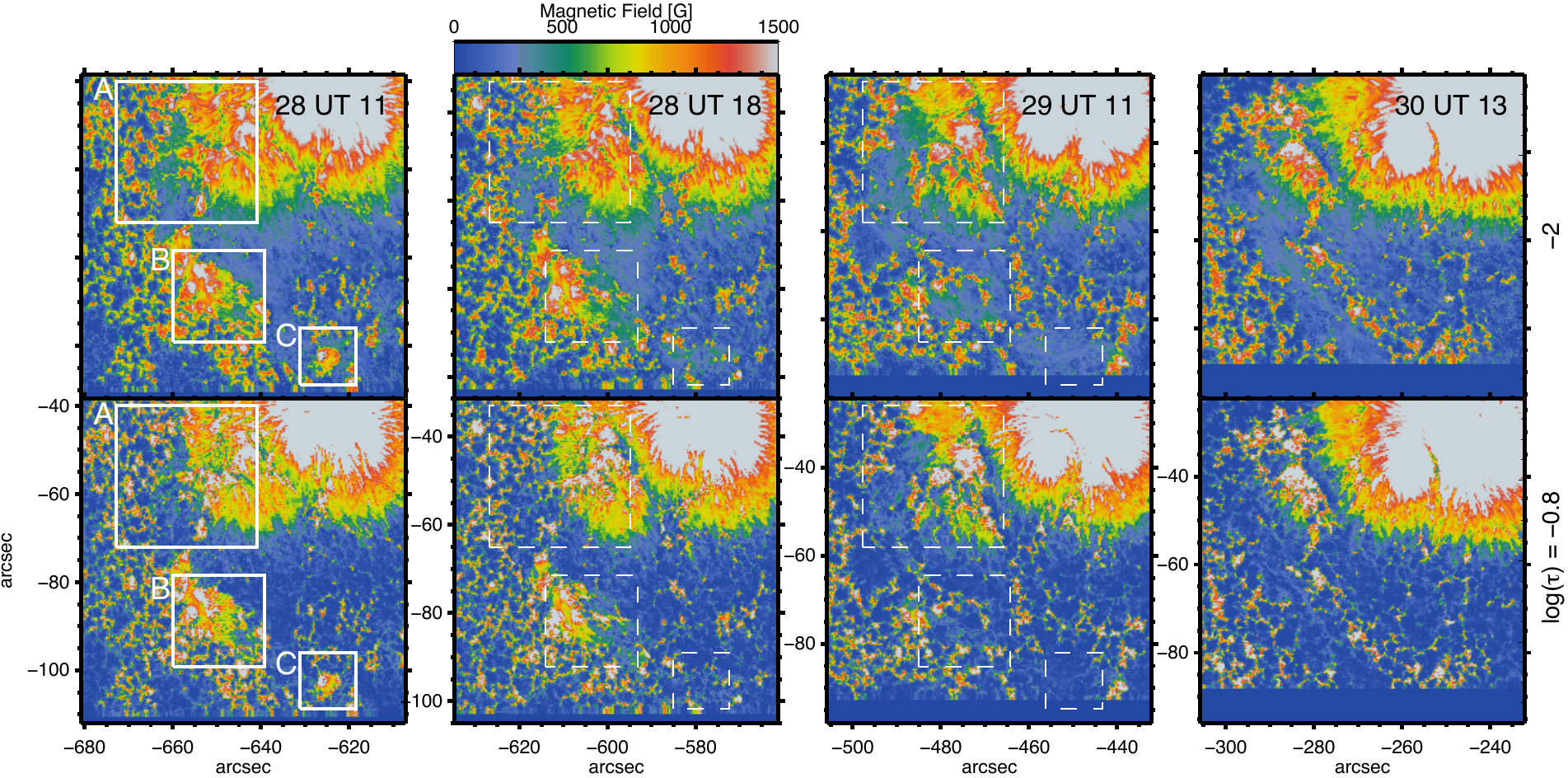}
        \caption{Magnetic field strengths obtained from the inversions. Each column corresponds to a separate SOT/SP scan, and the upper and lower rows indicate the field strengths at $\log(\tau)=-2$ $\&$ -0.8 respectively. The colour bar applies to all plots and the $xy$ axes indicate the distance to disc centre. The three $\emph{white}$ boxes are identical to those in Fig. \ref{Overviewevo}.}
         \label{MagOverviewevo}
         \end{figure*}

The evolution of the magnetic field strengths is shown in Fig. \ref{MagOverviewevo}. The two rows indicate the fields at the upper two $\log(\tau)$ nodes set during the inversion. At UT 11 on $28^{th}$ April regions $A$, $B$ and $C$ all contain kG fields across all three $\log(\tau)$ layers, which match the locations of reduced temperature seen in Fig. \ref{Overviewevo}. As the three regions evolve, the kG fields give way to a more diffuse hG field (light blue colour in Fig. \ref{MagOverviewevo}), which is prevalently situated in the upper node at $\log(\tau)=-2$. It is nestled between the MFCs aligned along the PIL and is suspended above the photosphere, since the corresponding lower $\log(\tau)$  layers are almost field free.

Figure \ref{MagOverviewevo} also reveals that regions $A$, $B$ and $C$ do not evolve simultaneously but separately. Region $C$ is the first to evolve  and by UT 18 on $28^{th}$ April (column 2 Fig. \ref{MagOverviewevo}) hosts a hG field suspended at $\log(\tau)=-2$. Regions $A$ $\&$ $B$ initially remain relatively unchanged, but eventually follow suit and also host their suspended hG field by UT 11 on $29^{th}$ April (column 3 Fig. \ref{MagOverviewevo}), some 17 hours later. Note that, at this time the hG fields of the three regions are still separated from each other by several kG MFCs (e.g. $-470X$, $-100Y$). The separate hG field regions continue to expand (see column 4 Fig. \ref{MagOverviewevo}) and by UT 13 on $30^{th}$ April the hG fields form a continuous channel.

\begin{figure*}
       \centering
        \includegraphics[width=18cm]{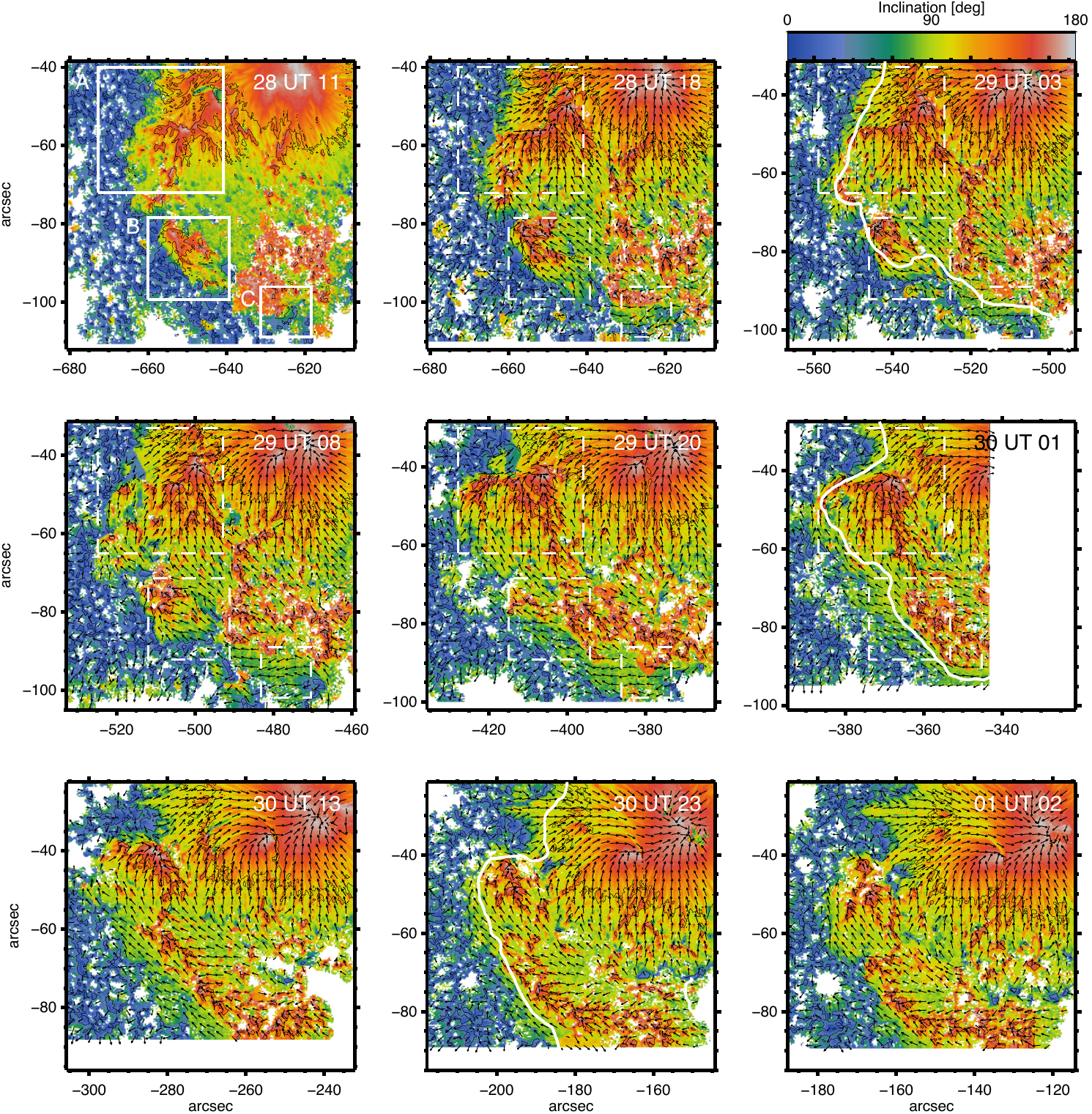}
        \caption{LOS inclinations at $\log(\tau)=-2$ of four SOT/SP scans. The $\emph{black}$ contour lines indicate kG magnetic fields at $\log(\tau)=-0.8$. The colour bar applies to all panels, where $\emph{blue}$ represents positive polarity fields and $\emph{red}$ the negative polarity. The arrows indicate the direction of the azimuthal magnetic field component at $\log(\tau)=-2$ and the $\emph{white}$ line follows the PIL. The $xy$ axes indicate the distance to disc centre. The three $\emph{white}$ boxes are identical to those in Fig. \ref{Overviewevo}.}
         \label{InclOverviewevo}
         \end{figure*}

Figure \ref{InclOverviewevo} displays the inclination of the magnetic fields at $\log(\tau)=-2$, while the black contour lines encompass kG MFCs. The presence of an extended plage region of opposite polarity to the sunspot can be discerned from this figure. Regions $A$, $B$ and $C$ are all placed on the PIL running through the AR (see $\emph{white}$ line at e.g. UT 01 on $30^{th}$ April in Fig. \ref{InclOverviewevo} for reference). Even though the magnetic fields in regions $A$, $B$ and $C$ evolve in time, the PIL is maintained throughout. Aligned along the PIL are MFCs, which is characteristic for most H$\alpha$ filament channels. The MFCs on the sunspot side are, furthermore, affected by the Sunspot's moat flow, which carries new MFCs towards the PIL \citep{okamoto2009,vargas2012}. The lower row panels in Fig. \ref{InclOverviewevo} illustrate this behaviour. Whilst the MFCs around the PIL contain predominantly vertical fields, the fields on the PIL itself are predominantly horizontal, especially once regions $A$, $B$ and $C$ have reduced in complexity. Furthermore, the azimuthal orientation of this field, displayed by the arrows in Fig. \ref{InclOverviewevo}, follows an $\emph{inverse}$ configuration (i.e. the field also points from negative to positive polarity) at e.g. UT 01 on $30^{th}$ April once the three regions have merged.
         
\begin{figure*}
       \centering
        \includegraphics[width=18cm]{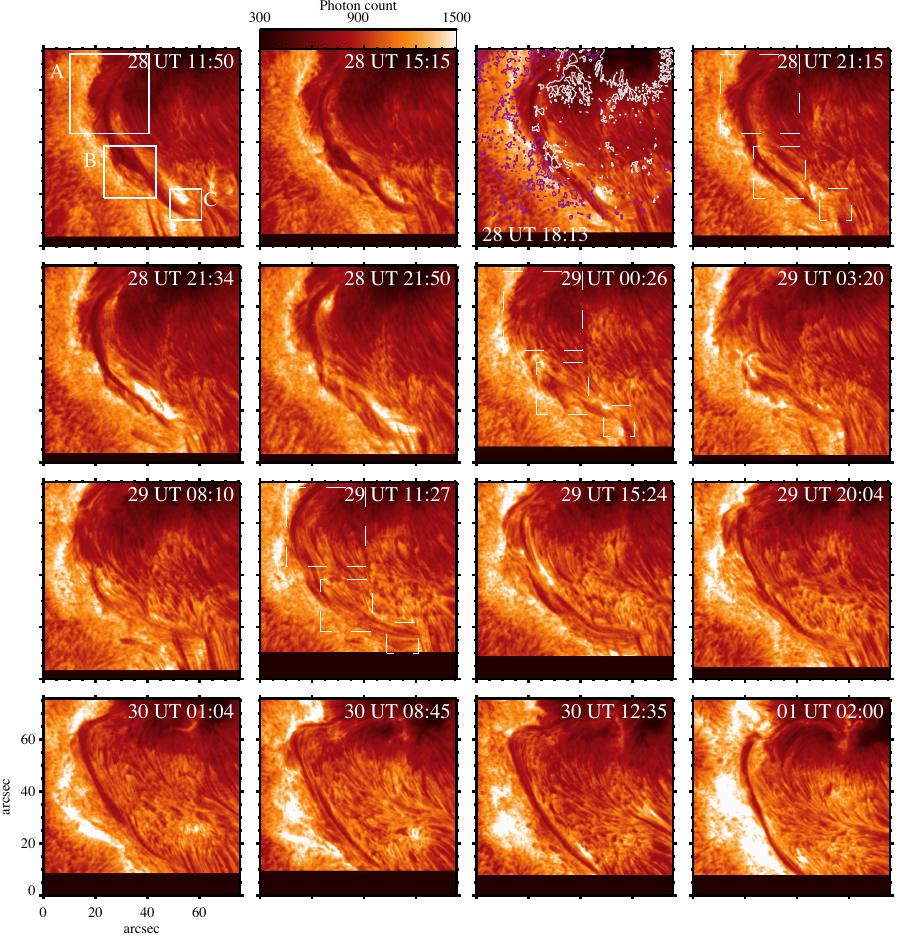}
        \caption{Sequence of SOT/NFI H$\alpha$ images. The colour bar applies to all plots and the $xy$ axes indicate the size of each image. The contour lines in the $3^{rd}$ frame the top row enclose kG magnetic fields at $\log(\tau)=-0.8$ obtained from the inversion of a SOT/SP scan. The $\emph{grey}$ contour lines in the third panel of the top row correspond to negative polarity and the $\emph{magenta}$ to positive polarity fields. The three $\emph{white}$ boxes are identical to those in Fig. \ref{Overviewevo}.}
         \label{HaOverviewevo}
         \end{figure*}

The horizontal hG field aligned along the PIL at UT 13 on $30^{th}$ April in Figs. \ref{MagOverviewevo} $\&$ \ref{InclOverviewevo} indicates the presence of an H$\alpha$ filament suspended above the photosphere. The SOT/NFI observations displayed in Fig. \ref{HaOverviewevo} not only support this assertion, but also reveal that the complex photospheric evolution seen in Figs. \ref{Overviewevo}, \ref{MagOverviewevo} $\&$ \ref{InclOverviewevo} is mirrored in the chromosphere. The observations show that a fully formed H$\alpha$ filament was already present above the PIL at UT 11:50 on $28^{th}$ April, which is near the start of the SOT/NFI observation run. Furthermore, the photospheric magnetic fields of regions $A$ $\&$ $B$ are situated directly underneath the H$\alpha$ filament, whereas region $C$ is slightly offset. The initial H$\alpha$ filament is stable until a series of small dynamic brightening events rooted at region $C$ (e.g.  $45X, 20Y$ at UT 21:34, 21:50 on $28^{th}$ April in Fig. \ref{HaOverviewevo}) begin to erode it. The events manifest themselves in the form of intense brightenings lasting no longer than two images (< 2 min), which are seen simultaneously in H$\alpha$ and Ca II H. They may be a manifestation of reconnection. By UT 00:22 on $29^{th}$ April the initial overlying H$\alpha$ filament has disappeared and given way to several disjointed dark fragments, located at regions $A$, $B$ and $C$. The azimuthal magnetic field component (see Fig. \ref{InclOverviewevo}) is always orientated in parallel with these fragments. The small fragments slowly elongate to coalesce into a continuous H$\alpha$ filament, by UT 20:04 on $29^{th}$ April, a development that is supported by the inversion results in Figs. \ref{MagOverviewevo} $\&$ \ref{InclOverviewevo}. The newly formed H$\alpha$ filament then remains stable until at least the $7^{th}$ of May, although a few sporadic brightening events do occur along the PIL.
  
The evolution of the PIL and the H$\alpha$ filaments located within it appears to be driven by the magnetic fields in regions $A$, $B$ and $C$. Region $C$ appears to be responsible for the destabilisation of the initial H$\alpha$ filament as well as providing one of the seeds from which the new H$\alpha$ filament is formed. Regions $A$ $\&$ $B$, while comparatively passive at first, also appear to be intimately involved in the creation of the new H$\alpha$ filament. Therefore, the three regions will be examined more closely.

\subsection{Region C}
Region $C$ is the southernmost region of the three and was outside the FOV of the SOT/BFI channels at all times. Due to the continuous northward drift of the FOV of all SOT channels, the majority of region $C$ was also outside the FOV of the SOT/SP and H$\alpha$ channels by UT 09 on $30^{th}$ April (see Figs. \ref{MagOverviewevo} $\&$ \ref{InclOverviewevo}). The first SOT/SP scan at UT 18 on $27^{th}$ April already shows the existence of the PIL, but contains no magnetic structure that can be linked to the pore situated at the heart of region $C$ at UT 11 on $28^{th}$ April (see Fig. \ref{Overviewevo}), suggesting that region $C$ is composed of freshly emerged flux. The second SOT/SP scan at UT 09 on $28^{th}$ April hints that region $C$ already contained a pore at that time, however, a tracking error meant that the region was not scanned in its entirety. Therefore, the emergence of the magnetic flux of region $C$ into the photosphere is unfortunately not covered by SOT.
 
\begin{figure}
       \centering
        \includegraphics[width=8cm]{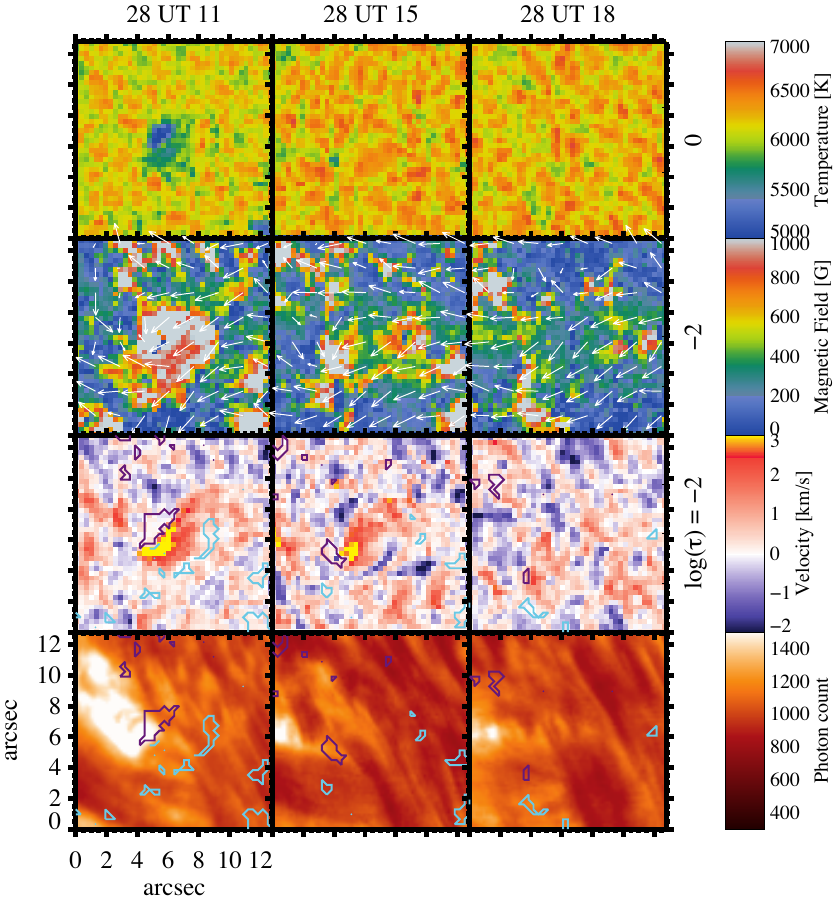}
        \caption{Sequence of inverted SOT/SP scans and H$\alpha$ images of region $C$. The $\emph{purple}$ contour lines correspond to negative polarity and the $\emph{turquoise}$ to positive polarity kG fields at $\log(\tau)=-0.8$. The arrows indicate the direction of the azimuthal magnetic field component at $\log(\tau)=-2$. The colour bars apply to all plots of their type and the $xy$ axes indicate the size of each plot. The $\log(\tau)$ layer of each plot is also indicated on the right of each row.}
         \label{regioncevo}
         \end{figure}

The subsequent evolution of region $C$ as seen by SOT/SP is displayed in Fig. \ref{regioncevo} and corresponds to SOT/SP scans 3 - 5 in Tab. \ref{SPtable}. At UT 11 on $28^{th}$ April the temperature at $\log(\tau)=0$ displays a dark pore at the centre of the FOV composed of kG magnetic fields in all $\log(\tau)$ layers. A somewhat less dark elongated structure, around $6X, 5Y$, is located immediately below it. The contour lines in Fig. \ref{regioncevo} reveal that this structure is located in between two kG magnetic field patches of opposite polarity and is itself host to horizontal fields of up to one kG. Over the two subsequent SOT/SP scans, covering 7 hours, the two opposite polarity kG patches drift apart from each other and reduce in size. The magnetic flux reduces by $50\%$ in all $\log(\tau)$ layers from $9 \times 10^{19}$ Mx to $4.5 \times 10^{19}$ Mx at $\log(\tau)=-0.8$. Simultaneously the horizontal fields connecting the kG patches, indicated by the $\emph{arrows}$ in Fig. \ref{regioncevo}, drop form $\sim$1 kG in all $\log(\tau)$ layers to $<100$ G in the lower two layers (see Fig. \ref{MagOverviewevo}) leaving a canopy field of $\sim$260 G at $\log(\tau)=-2$.  This canopy field persists in all subsequent observations (see Fig. \ref{MagOverviewevo}).

At UT 15 on $28^{th}$ April an upflow, co-spatial with the horizontal magnetic fields ($7X, 5Y$ in Fig. \ref{regioncevo}), can be seen between the two separating polarities in $\log(\tau)=-2$. The upflows reach a peak LOS velocity of 2 km/s. The configuration of upflows at the apex of rising magnetic fields together with downflows at their foot points supports the rising flux scenario. The dark fibrils seen in H$\alpha$ at UT 11 \& 15 in Fig. \ref{regioncevo} are generally orientated perpendicular to the photospheric fields, but beginning at UT 18 ($5X, 3Y$ in Fig. \ref{regioncevo}) at the location of the earlier photospheric upflows small fibrils connecting the opposite kG polarities and orientated in parallel with the photospheric field can be spotted. The photospheric fields and H$\alpha$ fibrils remain aligned for as long as region $C$ can be identified (see Figs. \ref{MagOverviewevo}, \ref{InclOverviewevo} \& \ref{HaOverviewevo}).

Furthermore, the azimuthal orientation of region $C$'s magnetic fields after they rose into the chromosphere already display an $\emph{inverse}$ configuration at $\log(\tau)=-2$ at UT 03 on $29{th}$ April in Fig. \ref{InclOverviewevo}. Together with the average field strength of 260 G and the co-spatial and aligned dark fibril seen at $55X, 15Y$ at the same time in Fig. \ref{HaOverviewevo}, reveal that region $C$ alone already displays the characteristics of the H$\alpha$ filament that forms out of regions $A$, $B$ and $C$ at UT 20:04 on $29^{th}$ April. Potentially, magnetic fields stemming from a small, compact region such as region $C$ may already be sufficient to form a small H$\alpha$ filament.

\subsection{Region B}
Region $B$ is the most striking and revealing region of the three. Its initial photospheric composition is characterised by pores and several elongated filamentary structures, which are reminiscent of orphan penumbrae. They are identifiable at UT 11 on $28^{th}$ April in Fig. \ref{Overviewevo}. All of these features are already fully formed in the first SOT/SP scan at UT 18 on $27^{th}$ April, which indicates that the emergence of region $B$'s magnetic flux into the photosphere occurred when the AR was still located on the far side of the Sun.
 
\begin{figure}
       \centering
        \includegraphics[width=8cm]{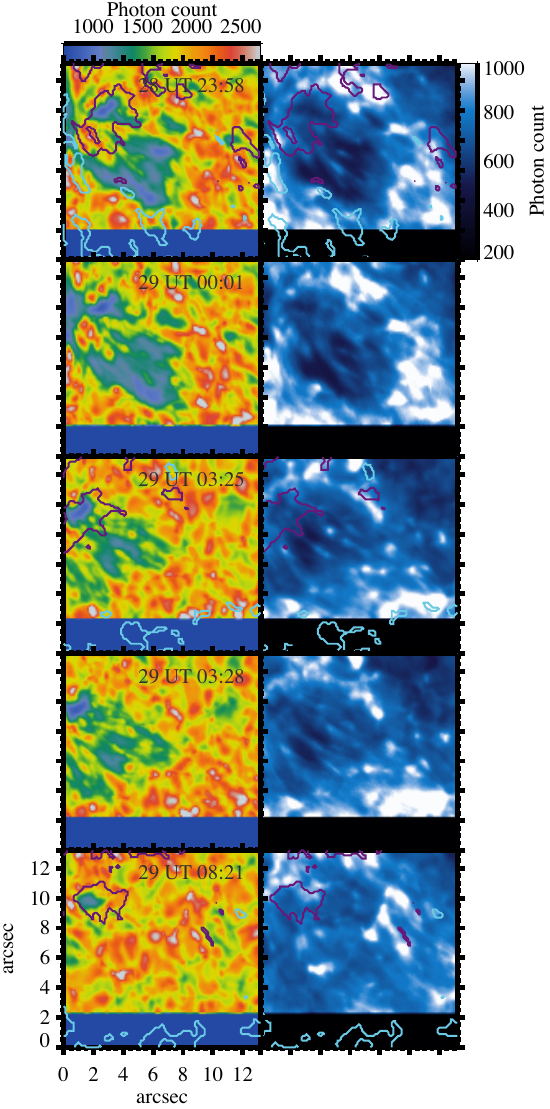}
        \caption{$\emph{Left}$: Sequence of SOT/BFI G-band images of region $B$. $\emph{Right}$: Sequence of co-spatial and co-temporal SOT/BFI Ca II H images. The contour lines enclose kG magnetic fields at $\log(\tau)=-0.8$ obtained from inversions of SOT/SP scans. The $\emph{purple}$ contour lines correspond to negative polarity and the $\emph{turquoise}$ to positive polarity fields. The colour bars apply to all their respective images and the $xy$ axes indicate the size of each image.}
         \label{caregionb}
         \end{figure}

From UT 11:39 on $28^{th}$ April onwards region $B$ was continuously monitored by SOT/NFI and the two SOT/BFI channels. The H$\alpha$ images in Fig. \ref{HaOverviewevo} indicate that the photospheric magnetic structures of region $B$ are located directly underneath the initial H$\alpha$ filament. Whilst the initial H$\alpha$ filament is in place the photospheric magnetic structures of region $B$ are very stable (see UT 11 \& 18 on $28^{th}$ April in Fig. \ref{MagOverviewevo}). Once the initial H$\alpha$ filament has disappeared the photospheric composition of region $B$ begins to change according to Figs. \ref{MagOverviewevo} \& \ref{HaOverviewevo}. The G-band and Ca II H observations of region $B$ support this assertion and a selection of images of both channels is displayed in Fig. \ref{caregionb}.
 
\begin{figure}
       \centering
        \includegraphics[width=8cm]{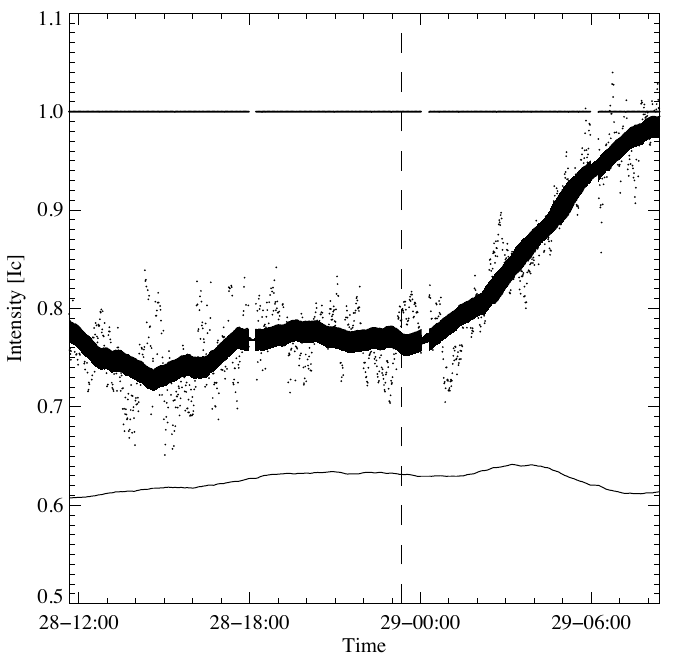}
        \caption{Evolution of the SOT/BFI G-band normalised mean intensity of the orphan penumbrae displayed in Fig. \ref{caregionb}. The $\emph{dots}$ are the individual measurements for each image and the $\emph{thick solid}$ line represents a 3 hour smoothing of these measurements. The continuous $\emph{thin solid}$ line indicates a 3 hour smoothing of normalised mean intensity measurements of sunspot penumbra. The vertical $\emph{dashed}$ line marks the time of the overlying H$\alpha$ filament's disappearance.}
         \label{regionbgbandevo}
         \end{figure}

The images in Fig. \ref{caregionb} give an account of the photospheric evolution of region $B$ beginning just after the disappearance of the initial overlying H$\alpha$ filament. At UT 23:58 on $28^{th}$ April both channels observed a pore and several orphan penumbrae connecting opposite polarities, derived from co-temporal SOT/SP observations. Over the following 8 hours the orphan penumbrae become brighter and shorter before they disappear entirely leaving only the pore. These orphan penumbrae were already identifiable in the first SOT/BFI images at UT 11:39 on $28^{th}$ April and the evolution of their G-band mean intensity is displayed in Fig. \ref{regionbgbandevo}. The $\emph{thick}$ $\emph{solid}$ curve represents a 3 hour smoothing of the individual mean intensity measurements and testifies to the initial stability of the orphan penumbrae followed by a gradual brightening. The $\emph{dashed}$ vertical line in the same figure marks the disappearance of the initial overlying H$\alpha$ filament. Therefore, Fig. \ref{regionbgbandevo} indicates that the gradual evolution and disappearance of the photospheric structure is triggered by a change in the overlying chromosphere.
 
The individual mean intensity measurements shown by the $\emph{dots}$ in Fig. \ref{regionbgbandevo} display a somewhat erratic behaviour, which is principally caused by the continuous drift of the FOV of SOT/BFI and causes the orphan penumbrae to be partially outside the FOV for short periods of time. The $\emph{solid}$ line centred at 0.62 I$_c$ in Fig. \ref{regionbgbandevo} represents a 3 hour smoothing of the sunspot's disc-side penumbra's mean intensity from the same G-band observations. As expected, the sunspot penumbra remains stable through the observations and is considerably darker than the orphan penumbra in region $B$. The brightness difference appears to be largely intrinsic, since no matter how the orphan penumbra in region $B$ is selected it is never as dark as the sunspot penumbra. The final selection of region $B$'s orphan penumbra was performed by drawing simple box around it, whose size is manually adjusted depending on the length and width of the structure. Each G-band image was also normalised to its own quiet Sun mean intensity to remove the limb darkening effect. A similar mean intensity evolution of the orphan penumbrae can also be observed for the Ca II H images over the same time period.
 
\begin{figure*}
       \centering
        \includegraphics[width=17cm]{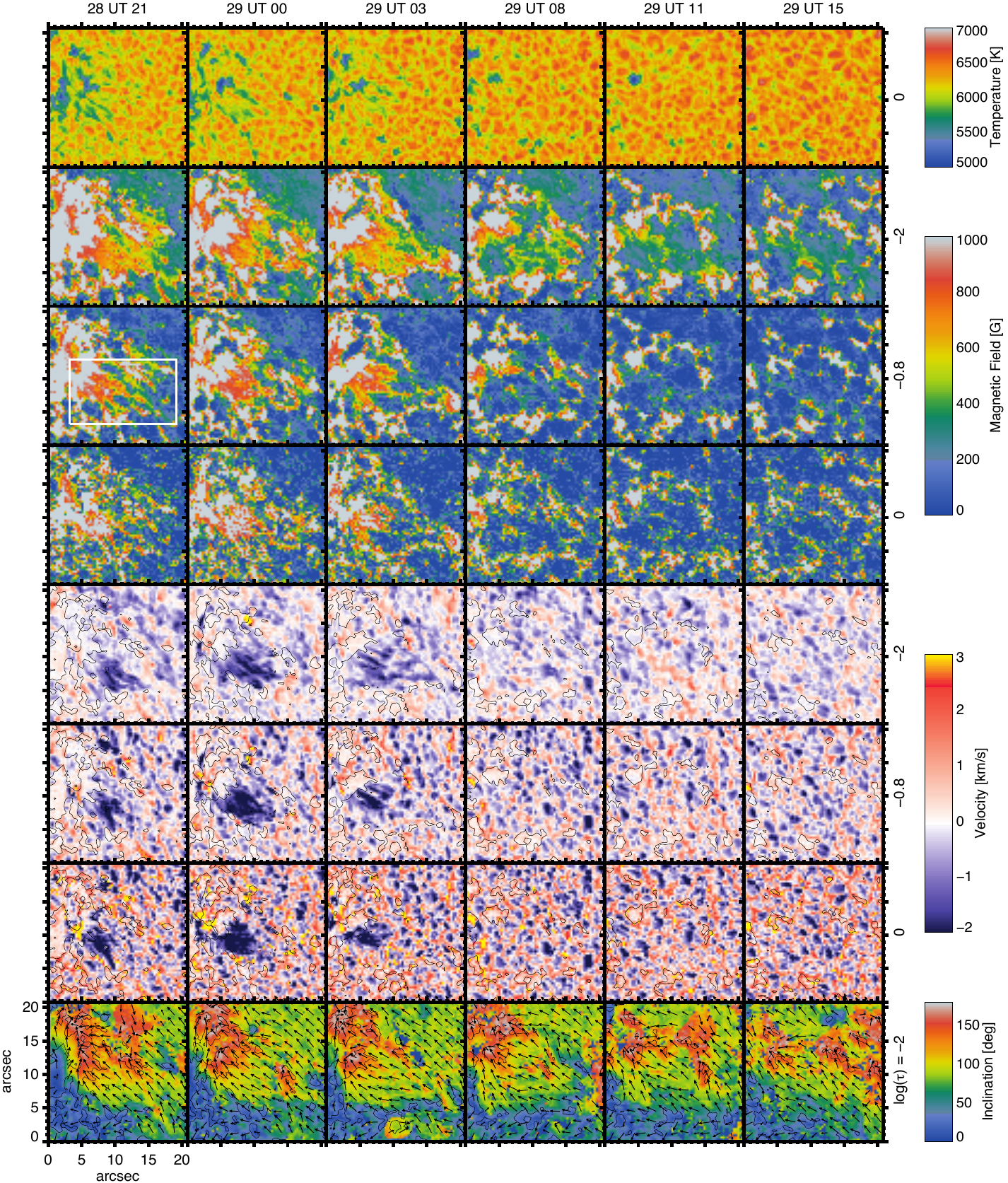}
        \caption{Sequence of inverted SOT/SP scans of region $B$. The $\emph{black}$ contour lines enclose kG magnetic fields at $\log(\tau)=-0.8$ and the arrows indicate the direction of the azimuthal magnetic field component at $\log(\tau)=-2$. The colour bars apply to all plots of their type and the $xy$ axes indicate the size of each plot. The $\log(\tau)$ layer of each plot is also indicated on the right of each row. The $\emph{white}$ box encloses rising magnetic fields.}
         \label{regionbevo}
         \end{figure*}

\begin{figure}
       \centering
        \includegraphics[width=8cm]{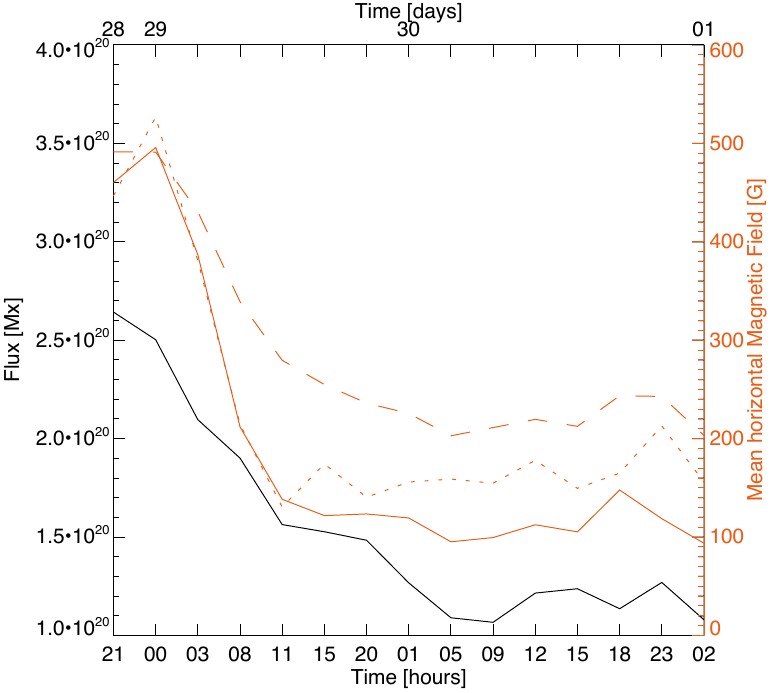}
        \caption{$\emph{Black}$: Change in magnetic flux at $\log(\tau)=-0.8$ in the region covered by Fig. \ref{regionbevo}. $\emph{Orange}$: Change in mean horizontal magnetic field strength within the $\emph{white}$ box in Fig. \ref{regionbevo}. The $\emph{dashed}$ line depicts $\log(\tau)=-2$, whereas the $\emph{solid}$ \& $\emph{dotted}$ lines correspond to $\log(\tau)=-0.8$ \& 0, respectively.}
         \label{regionbflux}
         \end{figure}

The SOT/SP inversions compiled in Fig. \ref{regionbevo} support and extend the impressions gained from Figs. \ref{caregionb} \& \ref{regionbgbandevo} and detail the evolution of region $B$ as obtained from the inversion of SOT/SP scans 6 - 11 in Tab. \ref{SPtable}. The temperature plots at $\log(\tau)=0$ in Fig. \ref{regionbevo} contain the same orphan penumbra around $8X, 8Y$. Just as in Fig. \ref{caregionb}, by UT 08 on $29^{th}$ April the orphan penumbra is no longer identifiable in the temperature plots in Fig. \ref{regionbevo}. Furthermore, the pore located at the head of the orphan penumbra has also disappeared by UT 15 on $29^{th}$ April. The number of kG features decreases sharply over the 18 hour period covered in Fig. \ref{regionbevo} with the consequent decrease in magnetic flux plotted in Fig. \ref{regionbflux}. Whilst the magnetic flux is close to constant over the first two time steps when the initial overlying H$\alpha$ filament was still intact, it rapidly decreases following the filament's disintegration. All the $\log(\tau)$ layers reveal a similar drop of $\sim$40 \% in magnetic flux over time between UT 00 and UT 15 on $29^{th}$ April.

The orphan penumbra, enclosed by a $\emph{white}$ box in Fig. \ref{regionbevo}, contain generally sub-kG fields at all times, except for a few pixels at $\log(\tau)=0$. For as long as the orphan penumbra is visible in the $\log(\tau)=0$ temperature plots, it contains sub-kG magnetic fields across all $\log(\tau)$ layers. In particular the fields at $\log(\tau)=0$ \& $-0.8$ closely follow the shape of the orphan penumbrae as seen in the temperature. By UT 03 on $29^{th}$ April the orphan penumbra has become somewhat shorter in length and this is reflected in the magnetic fields in the lower two $\log(\tau)$ layers. From UT 08 on $29^{th}$ April onwards the magnetic field has almost vacated the lower two $\log(\tau)$ layers and is predominantly situated in the upper $\log(\tau)$ layer. The change in the mean horizontal magnetic field strength of the pixels enclosed by the $\emph{white}$ box in Fig. \ref{regionbevo} across all $\log(\tau)$ layers is plotted in Fig. \ref{regionbflux} as well. The mean horizontal field strength at $\log(\tau)=-2$, represented by the $\emph{dashed}$ line in Fig. \ref{regionbflux}, settles at 260 G, forming a canopy over the lower two $\log(\tau)$ layers.

The magnetic field inclinations in Fig. \ref{regionbevo} demonstrate that the orphan penumbra is composed of predominantly horizontal fields, which connect opposite polarity kG MFCs. The negative polarity pore located at $5X,12Y$ in Fig. \ref{regionbevo} forms the head of the orphan penumbrae, whilst its tail is formed by several small MFCs located around $13X,5Y$. Once the overlying H$\alpha$ filament has disappeared the two opposite polarities move apart beginning at UT 00 on $29^{th}$ April. The separation of the polarities, also seen in Fig. \ref{caregionb}, suggests that the horizontal fields between them, originally forming the fields of the orphan penumbra, are rising into the chromosphere (as described in e.g. \citet{strous1996,lites1998,strous1999} \& \citet{centeno2012}). Unlike region $C$, the rise of the orphan penumbrae in region $B$ does not cause a series of intermittent brightenings visible in the H$\alpha$ and Ca II H data sets.

The orientation of region $B$'s orphan penumbra relative to the PIL (i.e. where $\Gamma\sim90^{\circ}$) undergoes a considerable evolution over time. From Fig. \ref{regionbevo} it is evident that the orphan penumbra, while it is identifiable in the $\log(\tau)=0$ temperature images (e.g. UT 21 on $28^{th}$ April), is aligned in parallel with the azimuthal magnetic field component. The azimuthal component in turn is orientated in a $\emph{normal}$ configuration with respect to the PIL (also see Fig. \ref{InclOverviewevo}). However, at UT 03 on $29{th}$ April and before, Figs. \ref{MagOverviewevo}, \ref{InclOverviewevo} \& \ref{HaOverviewevo} reveal that the orphan penumbra and its field is also already aligned in parallel to the initial H$\alpha$ filament as well as the H$\alpha$ filament that forms by UT 20:04 on $29^{th}$ April. The successive SOT/SP scans depicted in Fig. \ref{regionbevo} indicate the aforementioned loss of magnetic flux over time (see also Fig. \ref{regionbflux}), which causes the PIL in Fig. \ref{regionbevo} to rotate in particular from UT 11 to 15 on $29^{th}$ April. The azimuthal orientation of the orphan penumbra remains comparatively stable by comparison. Hence, the azimuthal orientation of the orphan penumbra relative to the PIL appears to rotate by up to $90^{\circ}$ over time. Note however, that the flux vacated the lower $\log(\tau)$ layers already several hours before the apparent rotation.

The LOS velocities in Fig. \ref{regionbevo} indicate fast flows in the orphan penumbra for as long as it is discernible in the $\log(\tau)=0$ temperature layer. At all times and across all $\log(\tau)$ layers the flows are co-spatial with the horizontal magnetic field of the orphan penumbra. The horizontal magnetic field at $\log(\tau)=-2$ is distributed over a greater area than at $\log(\tau)=0$, particularly at UT 03 on $29^{th}$ April, which is also reflected in the area covered by the high LOS velocities. Whilst at $\log(\tau)=0$ the flow often appears to stop short of reaching the opposite polarity kG MFC, forming the tail of the orphan penumbrae, it does reach the opposite MFC in the $\log(\tau)=-2$ layer. Therefore, the flow appears to be partially channelled above the photosphere during the structure's rise into the chromosphere, which may play a role in supplying mass to the forming H$\alpha$ filament. The flow stops abruptly in all the SOT/SP and SOT/BFI observations once the orphan penumbra's horizontal field has vacated the lower two $\log(\tau)$ layers.

\begin{table}
\caption{Comparison between region $B$'s orphan penumbra and sunspot penumbra at UT 00 $29^{th}$ April.}              
\label{comp}      
\centering                                      
\begin{tabular}{l c r r}          
\hline\hline                        
\bf{Parameters} & \bf{$-\log(\tau)$} & \bf{Region $\bf{B}$} & \bf{Penumbra}\\
\\  
\hline\hline                                   
    $T$ [K]                & 2.0    & 4810 & 4640\\      
                                & 0.8    & 5230 & 5060\\
                                & 0.0    & 6100 & 5910\\
\hline                      
    $B$ [G]                & 2.0    & 710 & 1160\\
                                & 0.8    & 710 & 1240\\
                                & 0.0    & 690 & 1320\\
\hline                      
 $\Gamma$ [$\circ$] & 2.0    & 100 & 110\\
                                & 0.8     & 98  & 104\\
                                & 0.0     & 97  & 103\\
\hline
   $\Phi$ [$\circ$]      & 2.0    &  154 & 151\\
                                 & 0.8    & 153 & 148\\
                                 & 0.0    & 148 & 148\\
\hline   
   $v$ [km/s]              & 2.0    & -0.8 & -0.3\\
                                 & 0.8    & -1.1 & -0.7\\
                                 & 0.0     & -0.8 & -1.5\\
\hline
   $\xi$ [km/s]            & 2.0    & 1.5 & 1.5\\
                                 & 0.8    & 2.1 & 2.1\\
                                 & 0.0    & 4.1 & 4.6\\
\hline\hline
                                             
\end{tabular}
\tablefoot{Each value represents a spatial average over the structures.}
\end{table}
Table \ref{comp} provides a quantitive comparison between region $B$'s orphan penumbra and a similarly orientated part of the sunspot's penumbra at UT 00 on $29^{th}$ April (e.g. see $-500X, -40Y$ at UT 03 on $29^{th}$ April in Figs. \ref{Overviewevo} $\&$ \ref{InclOverviewevo}). Each value represents a spatial average over each structure as identified by the temperature layer at $\log(\tau)=0$. Table \ref{comp} reveals several quantitative differences between region $B$ and the sunspot penumbra. Region $B$ is consistently hotter by $\sim$200 K across all $\log{\tau}$ layers, in line with the persistent continuum intensity difference between the two structures shown in Fig. \ref{regionbgbandevo}. The magnetic field strengths, too, are lower in region $B$ by $\sim$500 G. The LOS inclinations and azimuths are very similar, as expected from the selection criteria for both regions. The LOS velocities are of similar magnitudes, but the $\log(\tau)$ layer containing the highest average flows does differ. Region $B$'s orphan penumbra features consistently faster flows in the upper two $\log(\tau)$ layers when compared to the sunspot penumbra and, together with the lower magnetic field strengths, might indicate that it is geometrically located higher in the solar atmosphere than the sunspot penumbrae. Nonetheless, the clear spine/inter-spine structure of the orphan penumbra in region $B$, displayed in Fig. \ref{caregionb} and in all the inversion parameters listed in Fig. \ref{regionbevo} are in keeping with the $\emph{standard}$ penumbral filament described by \citet{tiwari2013}.

For as long as the orphan penumbra in region $B$ is identifiable in the G-band and Ca II H images (see Fig. \ref{caregionb}) both channels also reveal several bright spines separated by comparatively darker inter-spines in the orphan penumbrae e.g. around $5X,6Y$ at UT 23:58 on $28^{th}$ April in Fig. \ref{caregionb}. Along the spines small bright points can be made out, which travel continuously from the pore to the structure's tail. Bright points seen in the G-band images generally display the same behaviour except for bright points located within 3" of the pore and at the head of the orphan penumbra, which show an apparent counterflow directed towards the pore. In the Ca II H images these small bright points can be traced at all times and travel the entire way to the tail's opposite polarity patch, which is situated at the lower edge of the FOV and encompassed by the $\emph{turquoise}$ contour lines in Fig. \ref{caregionb}. By manually tracking 10 bright points they were found to have a typical plane of the sky velocity of 1.7 km/s in the Ca II H images at UT 03:25 on $29^{th}$ April (see Fig. \ref{caregionb}). Furthermore, the Ca II H observations support the suggestion made by Fig. \ref{regionbevo} that during the final ascent stages at UT 03 on $29{th}$ April part of the flow in the orphan penumbrae appears to be channelled above the photosphere.

Finally, an estimate of the rise speed of the orphan penumbra through the photosphere can be obtained by combining the rise time of $\sim$8 hours displayed in Fig. \ref{regionbgbandevo} and taking a value of $\sim$300 km for the thickness of the photosphere, which can be obtained from contribution functions of the 6302 $\AA$ line pair calculated by STOPRO. The resultant rise speed amounts to $10\pm5$ m/s. Therefore, the LOS velocities in the orphan penumbrae primarily depict a flow orientated in parallel to the solar surface akin to the Evershed flow \citep{evershed1909}. A similar conclusion about the flow direction can be gleaned by combining the bright point velocity in the Ca II H observations and the LOS velocity observations in Fig. \ref{regionbevo}.

\subsection{Region A}
Region $A$ is by far the most complex region of the three. The complexity is caused by the region's vicinity to the sunspot, the multitude of pores it harbours and the presence of a large orphan penumbra. The characteristic features of region $B$ can also be observed for region $A$, some of them in a more striking fashion.

       \begin{figure}
       \centering
        \includegraphics[width=8cm]{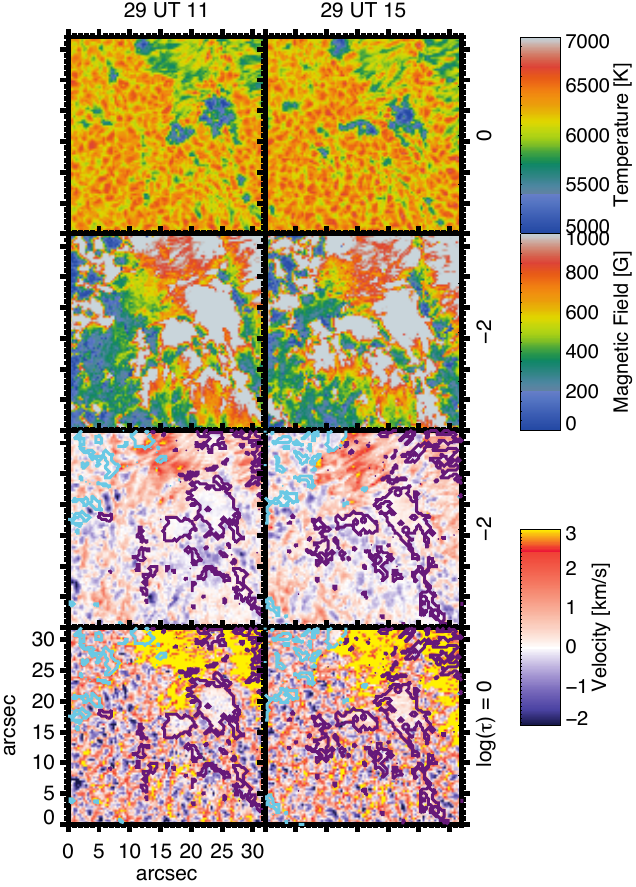}
        \caption{Sequence of inverted SOT/SP scans of region $A$. The $\emph{purple}$ contour lines correspond to negative polarity and the $\emph{turquoise}$ to positive polarity kG fields at $\log(\tau)=-0.8$. The colour bars apply to all plots of their type and the $xy$ axes indicate the size of each plot. The $\log(\tau)$ layer of each plot is also indicated on the right of each row.}
         \label{regionaevo}
         \end{figure}
The orphan penumbra in region $A$ also begins to evolve once the initial H$\alpha$ filament has disappeared, but it evolves somewhat slower than region $B$'s orphan penumbra.  Two SOT/SP scans of region $A$ revealing a similar evolutionary stage as region $B$ at UT 03 on $29^{th}$ April are presented in Fig. \ref{regionaevo}. They correspond to SOT/SP scans 10 and 11 in Tab. \ref{SPtable}. The orphan penumbra is located around pixel $20X,25Y$ and the edge of the sunspot penumbra around $30X,30Y$ in Fig. \ref{regionaevo}. The temperature and magnetic field contours reveal that part of the orphan penumbra in region $A$ is aligned in parallel to the H$\alpha$ filament channel. The magnetic fields at $\log(\tau)=-2$ extend continuously from the orphan penumbrae into the PIL, forming a hG field canopy (see Fig. \ref{MagOverviewevo}). The orphan penumbra, excluding the pores and MFCs, loses 30\% of its magnetic flux between UT 00 and UT 15 on $29^{th}$ April from $1.1 \times 10^{20}$ Mx to $8 \times 10^{19}$ Mx at $\log(\tau)=-0.8$. However, an exact measure of the lost magnetic flux due to the rise of the orphan penumbra is difficult to determine given that many MFCs move considerable distances in the area of interest. 
 
The orphan penumbra in Fig. \ref{regionaevo} hosts fast LOS velocities, the majority of which are akin to the Evershed flow in the sunspot penumbra also displayed in the figure. Unlike region $B$'s orphan penumbra, the sunspot penumbra and orphan penumbra in Fig. \ref{regionaevo} both display the fastest downflows at their tails across all $\log(\tau)$ layers. At $\log(\tau)=-2$ the downflows extend beyond the orphan penumbrae's $\log(\tau)=0$ temperature boundary and into the PIL. The extended outflow, located at $10X,20Y$, is traceable in both SOT/SP scans displayed in Fig. \ref{regionaevo} and represents a qualitatively similar case as the one seen at UT 03 on $29^{th}$ April in Fig. \ref{regionbevo}. Underneath the extended outflow a typical granulation pattern can be seen in the lower $\log(\tau)$ layers.
  
The extended outflow in region $A$ at UT 11 on $29^{th}$ April appears to return to the photosphere along two positive polarity MFCs situated at $5X,4Y$ and $8X,1Y$. Around the two MFCs at $\log(\tau)=-2$ the downflows are spatially extended, asymmetric and face the outflow of the orphan penumbra of region $A$. They are traceable in the lower two $\log(\tau)$ layers as well and reach a peak velocity of 6.6 km/s at $\log(\tau)=0$. Also, the two MFCs are $\sim$200 K hotter in the upper two $\log(\tau)$ layers when compared to other nearby MFCs. The Ca II H observations features persistent brightenings at the location of the MFCs. In Fig. \ref{regionaevo} and the Ca II H observations the two MFCs are seen to recede away from the orphan penumbra over time indicative of rising magnetic fields. The orphan penumbra's outflow would, therefore, cover a distance of $\sim$30'' and flow along the PIL. Both the extended outflow of the orphan penumbra is traceable in the Ca II H observations of region $A$ by the movement of small bright points. The extended outflows appear to be a characteristic property of ascending orphan penumbrae.

\section{Discussion}
The evidence presented in Sect. 3 allows us to surmise that the formation process of the H$\alpha$ filament formed by UT 20:04 on $29^{th}$ April 2007 in AR 10953 is driven by the rise of horizontal magnetic fields from the photosphere into the chromosphere. In this section we compare and contrast our findings to the literature.

The investigation by \citet{okamoto2008} employed Milne-Eddington inversions of SOT/SP data sets 10, 11, 12, 14, 17 and 20 from Tab. \ref{SPtable}, to infer the rise of a helical flux rope along the PIL of AR 10953. In particular, the authors focussed on a FOV encompassing region $B$ and parts of region $C$. According to the employed SOT/SP data sets the emergence is proposed to occur from UT 11 on $29^{th}$ April until UT 02 on $1^{st}$ May. However, the horizontal magnetic fields in both regions rose into the chromosphere before UT 11 on $29^{th}$ April, some 11 hours earlier for region $B$ and 24 hours for region $C$ (see Figs. \ref{regioncevo} \& \ref{regionbevo}). The H$\alpha$ filament, seeded by the horizontal magnetic fields of regions $A$, $B$ and $C$, reformed by UT 20:04 on $29^{th}$ April (see Fig. \ref{HaOverviewevo}), making the emergence time-line proposed by \citet{okamoto2008} irreconcilable with both the SOT/SP and SOT/NFI observations.
 
The helical flux rope emergence proposed by \citet{okamoto2008} is supposed to occur simultaneously along the entire PIL, whilst leaving no discernible sign in the continuum images. Flux emergences of order of 1 $\times$ $10^{19}$ Mx or more typically produce elongated granules \citep{centeno2012,vargas2014} or dark filamentary structures \citep{strous1996,lites1998,strous1999,kubo2003} visible in the continuum, which were not reported by \citet{okamoto2008} over the proposed time frame. Flux emergence even at the smallest scales \citep{centeno2007,danilovic2010sunrise} always entails an increase in magnetic flux, which is not supported by the SOT/SP observations in Sect. 3 nor by SoHO/MDI observations of AR 10953 \citep{vargas2012}. In addition, magneto-hydrodynamic (MHD) simulations of flux rope emergence reveal that the magnetic field emerges in a piecemeal fashion and at several locations, after it has been distorted by and subjected to the solar convection \citep{cheung2007,cheung2008,cheung2010,martinezsykora2008,stein2012}. MHD simulations, where a horizontal flux rope emerged along the entire length of the simulation box, still displayed an initial increase in magnetic flux and deformations in the granules \citep{yelles2009emer,mactaggart2010}. Regions $A$, $B$ and $C$ all contain structures visible in the continuum typical of flux emergence. They may represent the individual flux emergence sites of a larger connected flux system rising through the photosphere.
   
\citet{okamoto2008} observed a rotation in the magnetic field's azimuthal direction with respect to the PIL of almost $\sim$50$^{\circ}$ from UT 11 on $28^{th}$ April until UT 02 on $1^{st}$ May. The field appears to turn from an initially $\emph{normal}$ to finally an $\emph{inverse}$ configuration, even though the horizontal magnetic fields of both region $B$ and $C$ have already risen into the chromosphere by that time. The observed azimuth rotation is, therefore, a spurious effect brought about by the decay of magnetic flux in region $B$, which continued for some time after the horizontal field's ascend (see Fig. \ref{regionbflux}). The decay of flux locally effected a rotation of the PIL itself rather than in the azimuth of the magnetic field (see UT 11 $\&$ 15 on $29^{th}$ April in Fig. \ref{regionbevo}). Furthermore, the azimuthal orientation of region $C$'s horizontal magnetic fields, which ascended into the chromosphere some 24 hours before UT 11 on $28^{th}$ April (see Fig. \ref{MagOverviewevo} \& \ref{regioncevo}), are displayed in the lower right hand corner of Fig. 2's FOV in \citet{okamoto2008}. They exhibit an $\emph{inverse}$ configuration throughout the time-line used by \citet{okamoto2008}, which defies their proposed scenario.

The $\emph{sliding door mechanism}$, a widening followed by a narrowing of the PIL, was proposed by \citet{okamoto2008} as evidence and photospheric signature for a magnetic flux rope that rises through the photosphere and into the corona. It occurs at the same time as the azimuth rotation. Since our investigation uses the same SOT/SP data sets, the $\emph{sliding door mechanism}$ can also be seen in, for example, Fig. \ref{InclOverviewevo}. However, in light of the evidence presented in Sect. 3, the $\emph{sliding door mechanism}$ cannot be a sign of flux emergence. It appears to be a combination of flux decay (see Figs. \ref{InclOverviewevo}, \ref{regionbevo} \& \ref{regionbflux}), which caused a widening of the PIL, and the continuous moat flow around the sunspot (see Fig. \ref{InclOverviewevo} \& \citet{okamoto2009,vargas2012}), which slowly pushed new MFCs into the PIL, hence causing it to narrow. In the Ca II H observations the widening of the PIL is seen as a growing dark area (see Fig. 1 in \citet{okamoto2009}), in line with a decay of kG MFCs. The subsequent narrowing of the PIL manifests itself by a reduction of the dark area, by encroaching bright kG MFCs. The narrowing of the PIL (see Fig. 1 in \citet{okamoto2009}) is greatest at the sunspot-facing side of the PIL in keeping with a moat flow-aided replenishment of MFCs to the PIL.

The average photospheric magnetic field strength of 260 G at $\log(\tau)=-2$ in the filament channel at UT 20:04 on $29^{th}$ April (see Fig. \ref{HaOverviewevo}), when the new H$\alpha$ filament has formed, is consistent with previous such measurements \citep{lites2005,okamoto2008,lites2010,xu2012,sasso2014}. At several locations within the PIL higher fields strengths of up to 800 G at $\log(\tau)=-2$ can be observed, which appear to be somewhat short lived, since they can never be observed at the same place in consecutive SOT/SP scans. Similarly high magnetic field strengths in H$\alpha$ filaments have been previously reported both in chromosphere \citep{kuckein2009,sasso2011} and photosphere \citep{kuckein2012}.

The H$\alpha$ filament created by UT 20:04 on $29^{th}$ April coincides with horizontal magnetic fields at $\log{\tau}=-2$, whose azimuthal component is predominantly aligned in parallel with the PIL, but also appears to point from negative to positive magnetic polarity (see Fig. \ref{InclOverviewevo}). This highly non-potential configuration of the azimuthal magnetic field component is commonly referred to as an $\emph{inverse}$ configuration, which has also been noted by \citet{okamoto2008} for the same H$\alpha$ filament and has been observed in many other cases \citep[e.g.][]{lites2005,lopez2006,kuckein2012}. The $\emph{inverse}$ configuration has been commonly interpreted as the photospheric observable of a chromospheric horizontal flux rope \citep[e.g.][]{priest1989}.

\citet{kuckein2012} observed orphan penumbrae aligned in parallel to the H$\alpha$ filament channel hosting an H$\alpha$ filament at various times. The supersonic downflow velocities located at the edges of their orphan penumbrae \citep{kuckein2012vel}, appear to be akin to the high downflows seen in region $A$ at the tails of the orphan penumbra and sunspot penumbra (see Fig. \ref{regionaevo}). The orphan penumbrae are interpreted as parts of an extremely low lying filament trapped in the photosphere. Although their orphan penumbrae also subsequently disappear, it is unclear from the available observations, whether they played a similar role as the orphan penumbrae described in Sect. 3.

Region $C$ (see Fig. \ref{regioncevo}) displays properties akin to a rising loop system \citep{solanki2003,lagg2004,xu2010} during its initial rise into the chromosphere. The brightening events leading to the eventual disappearance of the initial H$\alpha$ filament (see Fig. \ref{HaOverviewevo}), suggests an interaction and possible reconfiguration of the pre-existing chromospheric fields and the rising fields \citep{manchester2004,archontis2013}. However, regions $B$ and $A$ do not show similar brightening events and instead the photospheric observations (see Figs. \ref{regionbevo} \& \ref{regionaevo}) indicate that the orphan penumbra wholly rises into the chromosphere.

The investigations by \citet{guglielmino2014} \& \citet{zuccarello2014} analysed orphan penumbrae similar to regions $A$ and $B$ and concluded that they are a manifestation of emerged $\Omega$ loops, which are trapped in the photosphere. The evolution of region $B$ in particular both supports and complements their findings. The orphan penumbra in region $B$ is stable for at least 12 hours, whilst located underneath the initial H$\alpha$ filament. Once the overlying H$\alpha$ filament and, presumably, most of its horizontal magnetic field have disappeared the orphan penumbra can no longer be maintained in the photosphere and begins to ascend (see Fig. \ref{regionbgbandevo}). Furthermore, the sunspot penumbra formation process studied by \citet{shimizu2012,rezaei2012,lim2013} \& \citet{jurcak2014spot} indicates that sunspot penumbra formation only sets in after a sufficient horizontal magnetic field has been established within the overlying chromosphere. We speculate that there is a similar need for strong overlying chromospheric fields to form and maintain orphan penumbrae.

\citet{jurcak2014} performed a comparison between orphan and sunspot penumbrae and found many qualitative similarities, but, like region $B$, the orphan penumbrae featured higher intensities and higher Evershed-like velocities (see Tab. \ref{comp}). The orphan penumbrae studied by \citet{jurcak2014} submerged over time unlike the orphan penumbrae in regions $A$ and $B$, which ascended into the chromosphere. Sub-photospheric processes are likely responsible for this divergent evolution.

\emph{Delta} sunspots are often associated with producing and hosting H$\alpha$ filaments \citep{tanaka1991,gaizauskas1994,leka1996,lites1997}. The penumbra between the two opposite polarity umbrae is usually aligned in parallel to the PIL. Their large size allows a detailed analysis of their horizontal fields even when only a spatial resolution of $1''$ is available. \citet{lites1995} observed the emergence and decay of a $\emph{delta}$ sunspot, which was accompanied by the formation of an H$\alpha$ filament. The observations were interpreted as an ascent of horizontal fields from the photosphere into the chromosphere, thereby forming an H$\alpha$ filament. The evolution of the whole system shares many analogies with regions $A$, $B$ and $C$ such as, penumbra aligned with the PIL, persistent high/supersonic LOS velocities in the penumbra \citep{martinez1994}, magnetic flux decay during and after the ascend of the horizontal fields, little flaring activity during the ascent, and a remaining H$\alpha$ filament after large photospheric structures have been largely reduced to plage MFCs. The qualitative similarities between the results in Sect. 3 and those of \citet{lites1995} suggest that our observations may represent a scaled down version of the same overall process \citep{rust1994}.

\citet{martin1990} proposed a series of six conditions, which would automatically lead to the creation of an H$\alpha$ filament. The H$\alpha$ filament formation described in Sect. 3 supports the conditions proposed by her, although the original conditions did not include rising/emerging flux. The AR features a PIL that exists throughout the entire observation runs and supports an overlying coronal arcade. The coronal arcade has not been shown explicitly here, but has been independently observed and extrapolated by \citet{derosa2009}. It has also been extrapolated by \citet{wheatland2009}. The magnetic fields within the PIL are transverse in the photosphere (see Fig. \ref{InclOverviewevo}) and the initial H$\alpha$ filament as well as the individual fragments, which precede the fully formed H$\alpha$ filament, are all aligned parallel to the PIL (see Fig. \ref{HaOverviewevo}), in accord with \citet{martin1990}. The two key proposed photospheric observables leading to H$\alpha$ filament creation, are a continuous advection of small MFCs into the PIL \citep{gaizauskas1997,gaizauskas2001} and their cancellation presented as flux decay \citep{lites1995}. Whilst the PIL in AR 10953 experiences a continuous advection of fresh MFCs into the PIL by the sunspot's moat flow \citep{okamoto2009,vargas2012}, the magnetic structures in regions $A$, $B$, and $C$ involved in the formation of the H$\alpha$ filament were already present before the start of the observation runs. The photospheric flux decay, however, is observed for all three regions and in keeping with \citet{martin1990}. Since the conditions proposed by \citet{martin1990} were based on low resolution ($1''-2''$) filtergram data and LOS magnetograms, we can add to the proposed conditions that the flux decay is, at least in this case, initially accompanied by rising horizontal magnetic fields. Two of the three additional criteria added by \citet{martin1998} can be verified in our observations as well. The SOT/NFI H$\alpha$ and SOT/BFI Ca II H observations show mass motions along the H$\alpha$ filament and $\emph{inverse}$ magnetic field configuration in the PIL in the SOT/SP scans suggests the existence of helicity in the H$\alpha$ filament. Barbs \citep{martin1992,bernasconi2005,lin2008} in H$\alpha$ were not detected in our observations.   

\section{Conclusion}
The polarity inversion line (PIL) of active region (AR) 10953 was observed using Hinode SOT/SP scans, SOT/BFI G-band and Ca II H and SOT/NFI H$\alpha$ observations covering a time span starting at UT 18:09 on $27^{th}$ April 2007 and lasting until UT 06:08 on $1^{st}$ May 2007.

The SOT/NFI H$\alpha$ observations revealed an initial H$\alpha$ filament over the PIL, which eventually disappeared by UT 00:22 on $29^{th}$ April 2007 following a flux emergence at designated region $C$. A new H$\alpha$ filament formed along the PIL by UT 20:04 on $29^{th}$ April 2007. The formation process was driven by the rise of horizontal magnetic fields from the photosphere into the chromosphere at three separate locations along the PIL labelled regions $A$, $B$ and $C$. Several dark fragments, co-spatial with the risen horizontal fields of regions $A$, $B$ and $C$, were detected in the SOT/NFI H$\alpha$ observations, which eventually expanded to form a new continuous H$\alpha$ filament. The rise of the horizontal fields from the photosphere into the chromosphere was largely complete by UT 11 on $29^{th}$ April 2007, some 9-24 hours before the full formation of the H$\alpha$ filament in the SOT/NFI H$\alpha$ observations.

The initial emergence of regions $A$, $B$ and $C$ into the photosphere could not be observed, since regions $A$ and $B$ had already emerged when AR 10953 appeared on the east limb and region $C$'s emergence was not seen due to observational gaps on the $27^{th}$ of April 2007. Regions $A$ and $B$ were situated directly underneath the initial overlying H$\alpha$ filament, which essentially trapped the regions' magnetic fields in the photosphere. The horizontal fields, whilst trapped in the photosphere, took on the appearance of orphan penumbrae exhibiting Evershed-like flows. Once the overlying H$\alpha$ filament had disappeared, the orphan penumbrae became unstable and rose into the chromosphere forming $\emph{seed}$ fragments for the new H$\alpha$ filament. Furthermore, the horizontal fields in the orphan penumbrae in all regions were already aligned along the same axis as both H$\alpha$ filaments whilst trapped in the photosphere.
  
The largely horizontal fields of the orphan penumbrae rise into the chromosphere $\sim$9-24 hours before the H$\alpha$ filament reconstitutes itself. Based on the results presented here, we propose that, at least for some H$\alpha$ filaments, orphan penumbrae, already aligned in parallel with the H$\alpha$ filament channel prior to H$\alpha$ filament formation, play a key role in producing it.

\begin{acknowledgements}
The authors thank an anonymous referee for the valuable comments on the manuscript. 
Hinode is a Japanese mission developed and launched by ISAS/JAXA, with NAOJ as domestic partner and NASA and STFC (UK) as international partners. It is operated by these agencies in co-operation with ESA and NSC (Norway).
This work has been partially supported by the BK21 plus program through the National Research Foundation (NRF) funded by the Ministry of Education of Korea.
\end{acknowledgements}

\bibliographystyle{aa}
\bibliography{/Users/davidbuehler/Documents/Latex/TheBib_copy}{} 

\end{document}